\documentclass[10pt,twocolumn,letterpaper]{article}

\usepackage{iccv}              

\usepackage{graphicx}
\usepackage{array}
\usepackage{tikz}

\definecolor{iccvblue}{rgb}{0.21,0.49,0.74}
\usepackage[pagebackref,breaklinks,colorlinks,allcolors=iccvblue]{hyperref}

\def\paperID{6588} 
\def\confName{ICCV}
\def\confYear{2025}

\title{Efficient and robust 3D blind harmonization for large domain gaps}

\author{Hwihun Jeong$^{1,2}$ \qquad Hayeon Lee$^{1}$ \qquad Se Young Chun$^{1,3}$ \qquad Jongho Lee$^{1}$\\
$^{1}$Department of ECE, $^{3}$INMC \& IPAI, Seoul National University\\
$^{2}$BWH, Harvard Medical School\\
{\tt\small \{hwihuni, hiyeon, sychun, jonghoyi\}@snu.ac.kr}}
\begin{document}
\maketitle
\begin{abstract}
Blind harmonization has emerged as a promising technique for MR image harmonization to achieve scale-invariant representations, requiring only target domain data (i.e., no source domain data necessary). However, existing methods face limitations such as inter-slice heterogeneity in 3D, moderate image quality, and limited performance for a large domain gap. To address these challenges, we introduce BlindHarmonyDiff, a novel blind 3D harmonization framework that leverages an edge-to-image model tailored specifically to harmonization. Our framework employs a 3D rectified flow trained on target domain images to reconstruct the original image from an edge map, then yielding a harmonized image from the edge of a source domain image. We propose multi-stride patch training for efficient 3D training and a refinement module for robust inference by suppressing hallucination. Extensive experiments demonstrate that BlindHarmonyDiff outperforms prior arts by harmonizing diverse source domain images to the target domain, achieving higher correspondence to the target domain characteristics. Downstream task-based quality assessments such as tissue segmentation and age prediction on diverse MR scanners further confirm the effectiveness of our approach and demonstrate the capability of our robust and generalizable blind harmonization.

\end{abstract}    
\section{Introduction}
\label{sec:intro}

Magnetic resonance imaging (MRI) is a widely used medical imaging technique that provides detailed information of soft tissues. Recently, deep learning has been applied to MRI for various tasks, including disease classification \cite{ADclass, PDclass, class3}, tumor segmentation \cite{seg1, seg2}, and solving inverse problems \cite{qsmnet,varnet}. However, the effectiveness of deep learning in MRI is often limited by a challenge: a domain gap in MRI data. Variations can occur in MR images due to differences in scanner vendors, settings, or protocols, even when imaging the same patient. This variability creates domain gaps and hinders the performance of a neural network trained on datasets with a different domain. To address these challenges, researchers have developed several harmonization methods to align source domain images (\ie, images with domain gaps) with the characteristics of the target domain. These methods include both traditional techniques \cite{shinohara2017volumetric, nyul1999standardizing, 
shinohara2014statistical, mirzaalian2016inter, COMBAT, fortin2017harmonization} and deep learning approaches \cite{deepharmony, cyclegan1, cyclegan2, disentangle1, taskbased1, taskbased2, torbati2023mispel, bashyam2022deep}.

While these harmonization methods have shown promising results particularly when using deep learning-based methods, they typically require both source and target domain data for network training, still limiting generalization to unseen source domains. To address this challenge, the concept of “blind harmonization” has recently been proposed \cite{jeong2023blindharmony}. This approach enables the training of a harmonization network using only target domain data while allowing generalization to unseen source domain images. For example, BlindHarmony \cite{jeong2023blindharmony} utilizes an unconditional flow model trained exclusively on target domain data and produces a harmonized image that maintains a correlation with the input source domain image. This technique does not require source domain data during training.

Despite this unique advantage, several limitations exist (Fig. \ref{fig:bhdconcept}a). First, the previously proposed methods \cite{jeong2023blindharmony,beizaee2024harmonizing} struggle to harmonize source domain images with a large domain gap. Second, the normalizing flow models primarily functioned as image density estimators rather than image generators, due to the inherent limitations in the expressiveness of normalizing flows. This constraint necessitates iterative processes for image generation, impeding the computationally efficient creation of high-quality harmonized images. Lastly, current blind harmonization models are restricted to 2D image generation, which can lead to inter-slice heterogeneity when applied to a 3D volume.


Recently, efforts \cite{song2023consistency, salimans2022progressive, lipman2022flow, liu2022flow} were made to reduce sampling times in the diffusion model, enabling practical application. Among these, rectified flow \cite{lipman2022flow} learns a transport map between two arbitrary distributions, streamlining the generation process while maintaining high image quality. Furthermore, a few studies \cite{chung2023solving, lee2023improving, khader2023denoising, bieder2023memory, friedrich2024deep, song2024diffusionblend} explore the training of diffusion models particularly for 3D image generation. However, training with large 3D volumes remains a challenge due to the limitations of computational resources, which typically confine processing to smaller 3D volumes (\eg, patch-based training).

In this paper, we propose BlindHarmonyDiff, a novel 3D blind harmonization framework based on the edge-to-image model that was introduced in a few previous edge-conditioned image generation methods  \cite{wang2019cannygan,chen2024contourdiff,zeng2024reliable}. Our approach leverages the observation that MR images of the same subject, despite varying contrasts due to different acquisition settings, maintain consistent edge information. Building on this insight, we develop an edge-to-image model trained for target domain data to reconstruct the original image from the edge map detected by an edge-detection method. For inference, we apply the same edge detection method to a source domain image and input the resulting edge map into our trained model to generate a harmonized image. Moreover, unlike previous edge-conditioned image generation methods, we enable 3D generation with advanced 3D patch training and suppress hallucination using a refinement module with domain knowledge. Our key contributions are:

\begin{enumerate}
    \item We present a novel 3D harmonization framework that utilizes an edge-to-image model designed for harmonization tasks.
    \item Multi-stride patch methodology is proposed for efficient 3D model training, enhancing the ability to capture and generate 3D volumetric structures.
    \item To mitigate hallucination and ensure robust inference, we propose a refinement module that enforces physical consistency between source and target domain images.
\end{enumerate}

\section{Related works}
\label{sec:rw}

\subsection{MRI harmonization}
Several studies have developed techniques for harmonizing MR images between source and target domains. Traditionally, these methods have utilized image-level post-processing techniques, such as histogram matching \cite{shinohara2014statistical,nyul2000new,nyul1999standardizing} and statistical normalization \cite{fortin2017harmonization,COMBAT,shinohara2017volumetric}. However, these conventional methods often struggle to capture subtle differences between images from distinct domains. The advent of deep learning has led to advancements in harmonization methods. One notable approach is DeepHarmony \cite{deepharmony}, which employs an end-to-end supervised learning framework to establish mappings between source and target domains. However, it necessitates large datasets of traveling subjects, which can be challenging to be acquired. To overcome this limitation, researchers have turned to CycleGAN-based style transfer networks \cite{cyclegan1,cyclegan2}, which can learn mappings between images from different domains without requiring paired data. Moreover, recent developments have introduced separate networks for contrast and structure \cite{disentangle1,disentangle2, cackowski2023imunity}, allowing for greater flexibility in harmonization tasks. Recently, blind harmonization methods \cite{jeong2023blindharmony,beizaee2024harmonizing} have been proposed to harmonize images from an unseen source domain by leveraging a normalizing flow model trained exclusively on a target domain dataset.

\subsection{Patch-based training for diffusion models}
Studies have aimed to reduce the high training costs of diffusion models through patch-based training. Patch-DM \cite{ding2023patched} is one of such methods with feature collage techniques in feature space to avoid boundary artifacts when generating large-size images. PatchDDM \cite{bieder2023memory} utilizes fixed-size patches with coordinate encoding to reduce memory consumption during training. Patch Diffusion \cite{wang2024patch} employs diverse patch sizes to capture cross-region dependencies at multiple scales, enhancing data efficiency, especially for smaller datasets.
Both PatchDDM and Patch Diffusion use pixel-level coordinates and full-size inference to avoid boundary artifacts. However, PatchDDM overlooks capturing global dependencies due to fixed patch sizes, and Patch Diffusion focuses on 2D training and does not reduce memory usage, limiting large 3D data processing. 



\section{Methods}
\label{sec:method}
\subsection{Harmonization model}

\begin{figure*}
\centering
      {\footnotesize
(a) Normalizing flow-based blind harmonization (\textbf{OLD})

  \includegraphics[width=0.97\textwidth]{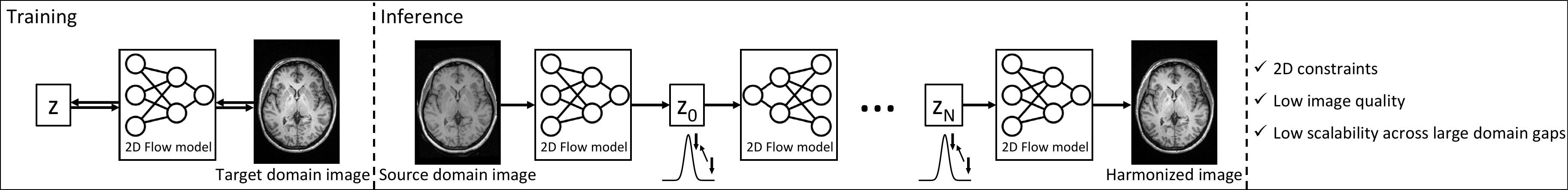}
  
(b) BlindHarmonyDiff (\textbf{NEW})

  \includegraphics[width=0.97\textwidth]{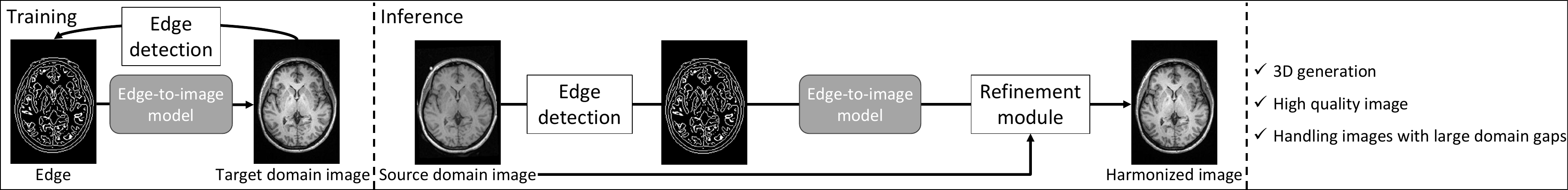}
  
(c) Multi-stride patch training

  \includegraphics[width=0.97\textwidth]{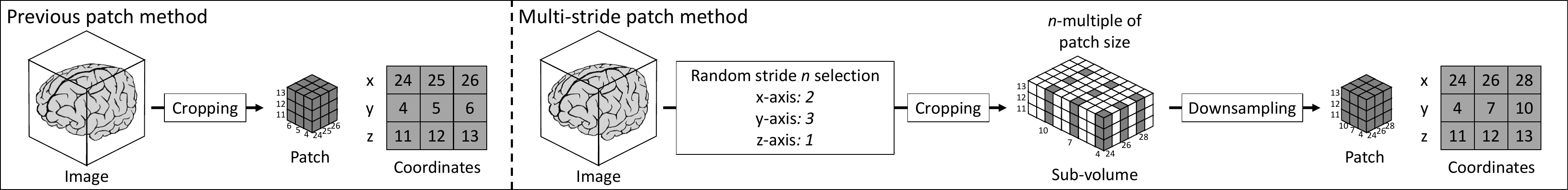}
  }
      \caption{(a) Previous blind harmonization methods have limitations such as 2D reconstruction, moderate image quality, and limited performance for a large domain gap. (b) BlindHarmonyDiff is a 3D blind harmonization method designed to overcome the limitations of the previous methods. BlindHarmonyDiff trains an edge-to-image model only with the target domain image and generates a harmonized image by inputting an edge map from the source domain image into the trained model. A refinement module then enhances the harmonized image to maintain a high correlation with the input source domain image. (c) An edge-to-image model is trained with a newly proposed multi-stride patch processing. As compared to a previous patch method, which simply creates a patch from neighboring voxels of an image, a multi-stride patch method randomly selects strides to construct a patch.}
\label{fig:bhdconcept}
\end{figure*}

The BlindHarmonyDiff framework is based on the assumption that MR images of the same subject acquired with different settings (\eg, vendor, scanner, parameters) exhibit varying contrasts but maintain a consistent edge \cite{jeong2023blindharmony}. Then, the construction of the target domain image ($x_{tar}$) and source domain image ($x_{src}$) of the same subjects can be formulated as below:
\begin{equation}
    x_{tar} = I(e,C_{tar}),  \,   x_{src} = I(e,C_{src}),
    \label{eq:imagemodel} 
\end{equation}
where $I$ is the image generation function from an edge and contrast, $e$ is a common edge, and $C_{tar}, C_{src}$ are the contrasts of the target and source domain image, respectively. This allows the edge detection method ($D$) to produce the same edge map from the target domain image ($x_{tar}$) and source domain image ($x_{src}$).
\begin{equation}
    D(x_{tar}) = D(x_{src}) = e. 
    \label{eq:edgedetect}
\end{equation}
Leveraging this assumption, we developed an edge-to-image model ($E2I$) trained to reconstruct an original image from its edge map, which was detected using an edge detection method applied to the target domain image (Fig. \ref{fig:bhdconcept}b).
\begin{equation}
    \theta^* =  \arg \min_\theta \mathcal{L}\left(E2I(D(x_{tar}),\theta),x_{tar}\right),
    \label{eq:e2itraing}
\end{equation}
where $\theta$ represents parameters of the edge-to-image model and $\mathcal{L}$ is a loss function for training. Using these facts, we can harmonize the source domain image to the target domain by detecting the edge of the source domain image to unlearn the contrast of the source domain and then reconstructing the contrast of the target domain with the trained edge-to-image model:
\begin{equation}
    x_{har} = E2I(D(x_{src}),\theta^*),
    \label{eq:bhdinfer}
\end{equation}
where $x_{har}$ is the resultant harmonized image (Fig. \ref{fig:bhdconcept}b).

\subsection{Edge detection method}
We employ Canny edge detection \cite{canny1986computational} as an edge detection method. To ensure consistent edge detection across different subjects and domains, we implement a subject-level Canny edge detection threshold determination approach. This approach maintains a uniform ratio of edge voxels to total image voxels. The algorithm iteratively adjusts the Canny edge detection threshold: if the ratio of edge voxels to total image voxels exceeds the predetermined value, the threshold is decremented. Through this process, we establish the Canny edge detection threshold that consistently yields 8\% edge voxels relative to the total image voxels.

\subsection{Patch-based rectified flow}
For training the edge-to-image model, we utilize coordinate-encoded patch-based diffusion model training strategies \cite{wang2024patch, bieder2023memory}, enabling efficient 3D training with limited resources and allowing inference for an entire 3D volume to avoid boundary artifacts. Our goal is to train a 3D patch-based edge-to-image model via rectified flow \cite{lipman2022flow}, which connects an image patch ($x_1$) and a Gaussian noise ($x_0$) on a straight trajectory:
\begin{equation}
dx_t = v(x_t, t) \, dt, \quad \text{where } x_t = t x_1 + (1-t) x_0,
\end{equation}
where $v$ represents the velocity field, parameterized as a neural network using the corresponding edge map patch ($e$). Spatial coordinates ($i, j, k$) are derived from the full images and normalized to range from -1 to 1, encoding positional information for each axis. The edge map patch ($e$) and positional coordinates ($i, j, k$) are used as conditions to control the rectified flow process. With the conditions ($e, i, j, k$), the velocity field ($v_\theta$) is optimized by minimizing the following loss function:
\begin{equation}
    \mathcal{L}= \mathbb{E}_{x_0\sim\mathcal{N}(0, 1)}  \left[ \int_0^1\lVert (x_1-x_0)-v_\theta (x_t,t;\,e,i,j,k) \rVert ^2  \, dt \right].
\end{equation}

\subsection{Multi-stride patch}
We propose a novel multi-stride patch approach (Fig. \ref{fig:bhdconcept}c), which introduces a new way to generate patches for patch-based training. Unlike previous methods that extract patches with a fixed size (\eg, $64^3$), our approach first randomly crops the whole 3D volume into a sub-volume with an n-multiple of the fixed patch size (\eg, 128$\times$192$\times$64, corresponding to 2$\times$, 3$\times$, and 1$\times$ of $64^3$, respectively). The sub-volume is then downsampled to the fixed patch size. To embed spatial context, we adjust and concatenate positional coordinates with the patches. This strategy maintains memory efficiency while effectively capturing cross-region dependencies, providing a coherent representation of global structures within the 3D data.

\subsection{Refinement module}
To address the potential hallucination issue inherent in diffusion models and enhance the reliability of harmonized images, we implement a refinement module applied to the edge-to-image model output. Because the difference between the source domain image and the target domain image primarily exists in the low-frequency region, the two images exhibit high correlation \cite{jeong2023blindharmony}. Leveraging this insight, we construct the refinement module by increasing the subject-wise normalized cross-correlation between the harmonized image and the input source domain image. Starting from the edge-to-image model output, the module performs gradient ascent, iteratively calculating the normalized cross-correlation between the source and generated images (step size = 0.02, the number of total iterations = 6; Fig. \ref{fig:bhdconcept}b).


\section{Experiments and results}
\label{sec:experiment}

\subsection{Dataset}\label{sec:dataset}
T$_1$-weighted images from the OASIS3 dataset \cite{lamontagne2019oasis} were used to train and evaluate the proposed framework. The OASIS3 dataset comprises images acquired using various scanners. We designated images scanned with the Siemens 3T Trio Tim scanner as the target domain. For the source domain, we employed four distinct datasets (1.5T Sonata, 3T BioGraph, 1.5T Vision, and 3T Magnetom Vida), each consisting of images acquired with different manufacturer models (details provided in Supplementary material \ref{sec:data}). All images underwent resampling to a resolution of 1.2$\times$1.2$\times$1.25 $\mathrm{mm}^3$. We applied image normalization using subject-level percentile-based min-max normalization, considering the 1st to 99th percentiles.

\subsection{Network training}
The 3D edge-to-image model was trained on cubic patches of size $64^3$ with a batch size of 8. The learning rate was set to 0.00005, and the Adam optimizer was employed. The training was conducted for 150K steps. The model architecture was based on a U-Net \cite{ronneberger2015u} with five input channels: one for Gaussian noise, one for the grayscale input, and three for the positional encoding of 3D coordinates (See Supplementary material \ref{sec:implementation} for detailed information).


\begin{figure}[t]
    \centering
    \includegraphics[width=0.8\linewidth]{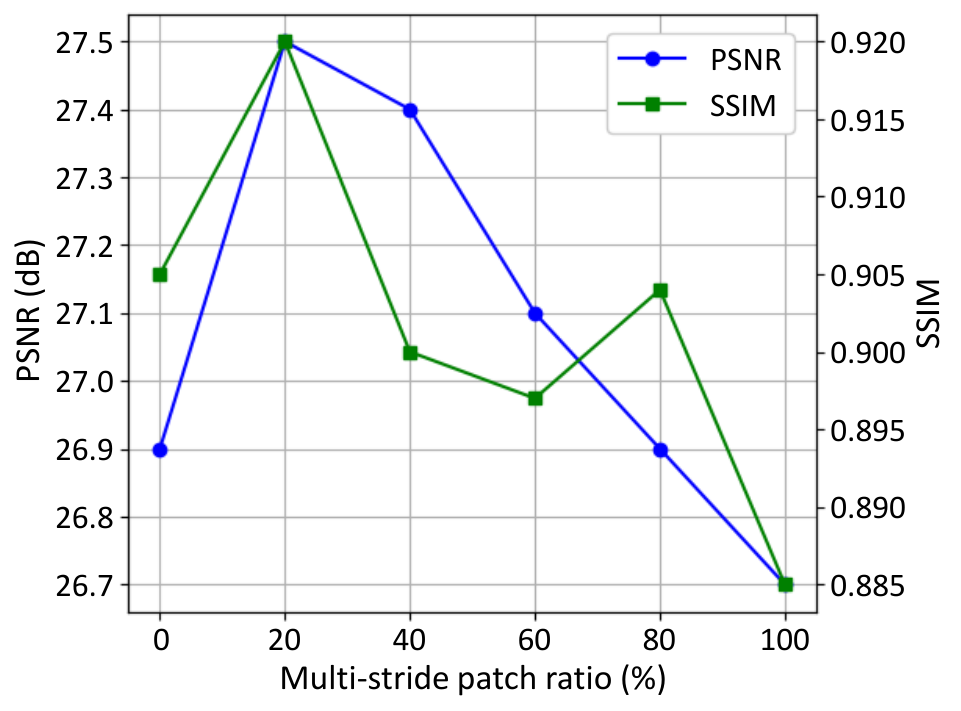}
    \caption{PSNR and SSIM across different multi-stride patch ratios. The ratio of multi-stride patches used in training ranged from 0\% to 100\% at 20\% intervals.}
    \label{fig:multi_stride_ratio}
\end{figure}

\begin{figure*}
\setlength{\tabcolsep}{0pt}
\renewcommand{\arraystretch}{0}
\centering
   {\scriptsize
\begin{tabular}{>{\centering\arraybackslash}m{0.05\textwidth}>{\centering\arraybackslash}m{0.118\textwidth}>{\centering\arraybackslash}m{0.118\textwidth}>{\centering\arraybackslash}m{0.118\textwidth}>{\centering\arraybackslash}m{0.118\textwidth}>{\centering\arraybackslash}m{0.118\textwidth}>{\centering\arraybackslash}m{0.118\textwidth}>{\centering\arraybackslash}m{0.118\textwidth}>{\centering\arraybackslash}m{0.118\textwidth}}
&Source domain & Target domain & BlindHarmonyDiff (ours) & BlindHarmony \cite{jeong2023blindharmony} & SSIMH  \cite{guan2022fast}& Histogram Matching  & Style transfer \cite{cyclegan1,cyclegan2} & DeepHarmony \cite{deepharmony}\\

          Sonata 1.5T&
         \includegraphics[width=0.118\textwidth]{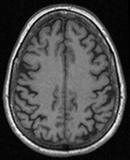}&
         \includegraphics[width=0.118\textwidth]{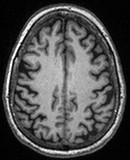}&
         \includegraphics[width=0.118\textwidth]{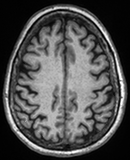}&
         \includegraphics[width=0.118\textwidth]{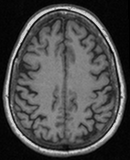}&
         \includegraphics[width=0.118\textwidth]{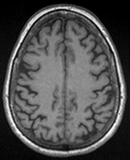}&
         \includegraphics[width=0.118\textwidth]{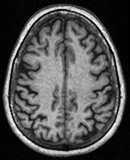}&
         \includegraphics[width=0.118\textwidth]{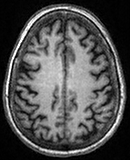}&
         \includegraphics[width=0.118\textwidth]{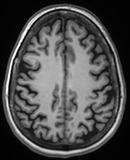}\\
         
        Vision 1.5T&
         \includegraphics[width=0.118\textwidth]{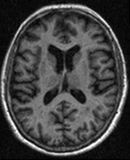}&
         \includegraphics[width=0.118\textwidth]{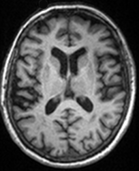}&
         \includegraphics[width=0.118\textwidth]{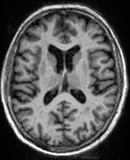}&
         \includegraphics[width=0.118\textwidth]{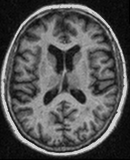}&
         \includegraphics[width=0.118\textwidth]{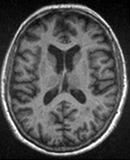}&
         \includegraphics[width=0.118\textwidth]{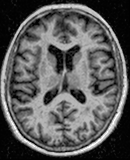}&
         \includegraphics[width=0.118\textwidth]{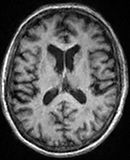}&
         \includegraphics[width=0.118\textwidth]{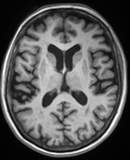}\\
         
     \end{tabular}
     }
     \caption{Visual comparison of harmonization results on traveling subjects, demonstrating the effectiveness of BlindHarmonyDiff against the other blind harmonization methods.}
   \label{fig:imgeval}
 \end{figure*}

\begin{table*}[]
\centering
      {\footnotesize
   \begin{tabular}{ccccccccccc}
   \hline
                                            &       & \multicolumn{2}{c}{Sonata 1.5T}     & \multicolumn{2}{c}{BioGraph 3T}     & \multicolumn{2}{c}{Vision 1.5T}     & \multicolumn{2}{c}{Magnetom Vida 3T} \\ 
                                            & Blind & SSIM($\uparrow$) & PSNR($\uparrow$) & SSIM($\uparrow$) & PSNR($\uparrow$) & SSIM($\uparrow$) & PSNR($\uparrow$) & SSIM($\uparrow$)  & PSNR($\uparrow$) \\ \hline \hline
   Source domain                            &      & 0.810          & 18.6               & 0.842          & 20.0               & 0.601          & 14.5               & 0.816          & 18.2                \\ \hline
   BlindHarmonyDiff (ours)                  & O    & \textbf{0.908} & \textbf{23.4}      & \textbf{0.862} & \textbf{23.3}      & \textbf{0.697} & \textbf{19.3}      & \textbf{0.851} & \textbf{21.3}       \\
   BlindHarmony \cite{jeong2023blindharmony} & O    & 0.802          & 18.9               & 0.850          & 21.3               & 0.665          & 17.2               & 0.822          & 19.0                \\
   SSIMH \cite{guan2022fast}                 & O    & 0.817          & 19.8               & 0.855          & 22.3               & 0.656          & 17.7               & 0.836          & 20.3                \\ 
   Histogram matching                       & O    & 0.894          & 21.8               & 0.857          & 22.9               & 0.678          & 18.6               & 0.844          & 20.7                \\ \hline
   Style transfer \cite{cyclegan1,cyclegan2} & X    & 0.895          & 24.3               & 0.864          & 23.5               & 0.667          & 19.2               & 0.855          & 22.7                \\
   DeepHarmony \cite{deepharmony}           & X    & 0.932          & 25.2               & 0.909          & 25.4               & 0.897          & 25.0               & 0.881          & 21.3                \\ \hline
   \end{tabular}}
   \caption{Quantitative assessment using PSNR and SSIM metrics, calculated relative to the target domain images. Results highlight the superior performance of BlindHarmonyDiff, reporting the best scores across the scanners.}
   \label{table:imagemetric}
\end{table*}

\subsection{Patch training for edge-to-image model}
We implemented a multi-stride patch training approach for the 3D edge-to-image model. Multi-stride patches were generated by efficiently downsampling sub-volumes to a fixed patch size via indexing, avoiding computationally expensive methods such as linear interpolation. During training, multi-stride patches (sampled with strides greater than 1) were used together with simple patches (sampled at a stride of 1), enabling the model to effectively learn both local and global features.
To determine the optimal ratio of multi-stride patches, we progressively increased their proportion from 0\% to 100\% during training. The performance of the edge-to-image model was evaluated across different ratios of multi-stride to simple patches using the target domain images (see Section \ref{sec:dataset}). PSNR and SSIM metrics were calculated relative to the original image to assess performance.

Our experimental results (Fig. \ref{fig:multi_stride_ratio}) demonstrate that the 20-80 split between multi-stride and simple patches produces the best PSNR and SSIM values. When compared to the simple patch training (\ie, 0\% in Fig. \ref{fig:multi_stride_ratio}), PSNR increased from 26.9 dB to 27.5 dB, and SSIM improved from 0.905 to 0.920. Multi-stride patch effectively balances local and global feature learning, leading to improved performance compared to simple patch training. The results hereafter are obtained with the edge-to-image model trained with a 20-80 split.

\subsection{Image level evaluation}
To evaluate the performance of harmonization, we utilized traveling subjects with data acquired across multiple scanners. Images from four source domains were registered to the target domain images using FSL FLIRT \cite{jenkinson2002improved}. The comparative analysis encompassed DeepHarmony \cite{deepharmony}, paired style transfer \cite{cyclegan1,cyclegan2}, and several blind harmonization techniques including normalizing flow-based blind harmonization (BlindHarmony) \cite{jeong2023blindharmony}, spectrum swapping-based image-level MRI harmonization (SSIMH) \cite{guan2022fast}, and histogram matching. For SSIMH implementation, we averaged the whole image of target domain images and replaced the low-frequency components of source domain images with the averaged target domain images using a 3D discrete cosine transform. Histogram matching was performed by storing the histogram of target domain images and matching the source domain image histogram to the target domain histogram at the subject level.

Figure \ref{fig:imgeval} demonstrates the harmonization of the source domain image (column 1) to the target domain (column 2) using various harmonization methods. BlindHarmonDiff (column 3) successfully harmonizes the source domain image, generating results with contrast similar to the target domain images. Compared to other blind harmonization methods (columns 4-6), BlindHarmonyDiff exhibits superior harmonization performance. While style transfer and DeepHarmony may demonstrate powerful harmonization capabilities, they are not blind harmonization methods, as their networks are trained on specific source domains.

Quantitative evaluation involves the calculation of PSNR and SSIM between harmonized and target domain images. The brain mask was extracted with FSL BET \cite{smith2002fast} and utilized for metric calculation. Table \ref{table:imagemetric} presents PSNR and SSIM values between the target domain image and harmonized images produced by each method. The ``Source domain'' row represents the source domain image without harmonization. Results demonstrate improved PSNR and SSIM metrics after applying BlindHarmonyDiff, with values comparable to those of style transfer methods trained for each source domain (average PSNR increase from 18.4 dB to 22.2 dB). Our method outperforms other blind harmonization techniques, including BlindHarmony, Histogram matching, and SSIMH, in both PSNR and SSIM metrics.
 
Normalizing flow-based blind harmonization exhibits lower PSNR and SSIM values compared to histogram matching and SSIMH. This discrepancy may be attributed to the differences in processing approaches: SSIMH and histogram matching were computed at the subject level using 3D processing, whereas BlindHarmony operated at the slice level with 2D processing. In the 2D case, BlindHarmony outperforms other blind harmonization methods, which is consistent with the previous results. (See Supplementary material \ref{sec:additional})

\begin{figure*}
\setlength{\tabcolsep}{0pt}
\renewcommand{\arraystretch}{0}
\centering
   {\scriptsize
\begin{tabular}{>{\centering\arraybackslash}m{0.05\textwidth}>{\centering\arraybackslash}m{0.105\textwidth}>{\centering\arraybackslash}m{0.105\textwidth}>{\centering\arraybackslash}m{0.105\textwidth}>{\centering\arraybackslash}m{0.105\textwidth}>{\centering\arraybackslash}m{0.105\textwidth}>{\centering\arraybackslash}m{0.105\textwidth}>{\centering\arraybackslash}m{0.105\textwidth}>{\centering\arraybackslash}m{0.105\textwidth}>{\centering\arraybackslash}m{0.105\textwidth}}
& Source domain image &No harmonization & Label & BlindHarmonyDiff (ours) & BlindHarmony \cite{jeong2023blindharmony}   & SSIMH \cite{guan2022fast}& Histogram Matching  & Style transfer \cite{cyclegan1,cyclegan2} & DeepHarmony \cite{deepharmony}\\
         Sonata 1.5T&
         \includegraphics[width=0.105\textwidth]{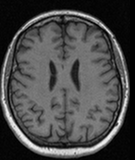}&
         \includegraphics[width=0.105\textwidth]{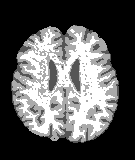}&
         \includegraphics[width=0.105\textwidth]{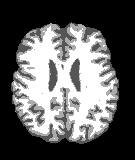}&
         \includegraphics[width=0.105\textwidth]{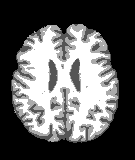}&
         \includegraphics[width=0.105\textwidth]{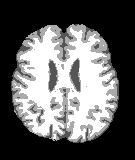}&
         \includegraphics[width=0.105\textwidth]{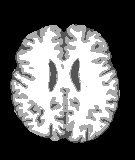}&
         \includegraphics[width=0.105\textwidth]{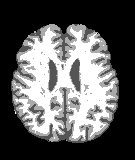}&
         \includegraphics[width=0.105\textwidth]{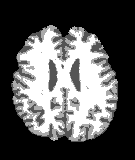}&
         \includegraphics[width=0.105\textwidth]{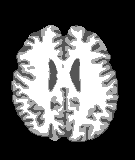}\\
         
         
        Vision 1.5T&
         \includegraphics[width=0.105\textwidth]{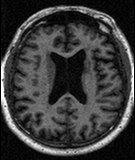}&
         \includegraphics[width=0.105\textwidth]{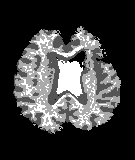}&
         \includegraphics[width=0.105\textwidth]{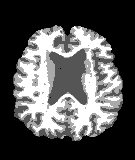}&
         \includegraphics[width=0.105\textwidth]{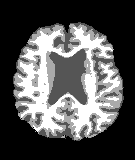}&
         \includegraphics[width=0.105\textwidth]{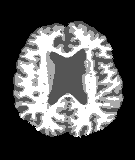}&
         \includegraphics[width=0.105\textwidth]{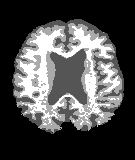}&
         \includegraphics[width=0.105\textwidth]{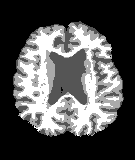}&
         \includegraphics[width=0.105\textwidth]{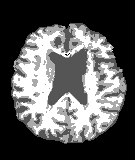}&
         \includegraphics[width=0.105\textwidth]{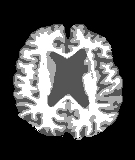}\\
         
     \end{tabular}
     }
     \caption{Tissue segmentation results for the various harmonization methods, with the segmentation network trained on the target domain data.  Each column represents a different harmonization approach applied to the input images.}
   \label{fig:downstream}
 \end{figure*}

\begin{table*}[]
\centering
      {\footnotesize
   \begin{tabular}{cc|>{\centering\arraybackslash}p{1cm}>{\centering\arraybackslash}p{1cm}>{\centering\arraybackslash}p{1cm}>{\centering\arraybackslash}p{1cm}|>{\centering\arraybackslash}p{1cm}>{\centering\arraybackslash}p{1cm}>{\centering\arraybackslash}p{1cm}>{\centering\arraybackslash}p{1cm}}
   \hline
                                            &           & \multicolumn{4}{c|}{Tissue segmentation: Dice($\uparrow$)}        & \multicolumn{4}{c}{Age prediction: MAE($\downarrow$)}\\ 
                                            & Blind     & Sonata        & BioGraph      & Vision        &  Vida             & Sonata        & BioGraph      & Vision        & Vida \\ \hline \hline
   No harmonization                         &           & 0.801         & 0.870         & 0.656         & 0.837             & 6.31          & 5.02          & 15.4          & 4.78          \\  \hline
   BlindHarmonyDiff (ours)                  & O         & \textbf{0.879}& \textbf{0.872}& \textbf{0.806}& \textbf{0.846}    & \textbf{5.60} & \textbf{4.91} & \textbf{6.94} & \textbf{4.63} \\
   BlindHarmony \cite{jeong2023blindharmony}& O         & 0.778         & 0.846         & 0.752         & 0.825             & 5.80          & 4.96          & 12.6          & 5.39          \\
   SSIMH \cite{guan2022fast}                & O         & 0.755         & 0.866         & 0.748         & 0.837             & 6.65          & 5.17          & 13.2          & 5.18          \\ 
   Histogram matching                       & O         & 0.869         & 0.865         & 0.793         & 0.835             & 6.19          & 5.08          & 10.5          & 5.34          \\  \hline
   Style transfer \cite{cyclegan1,cyclegan2}& X         & 0.807         & 0.851         & 0.697         & 0.804             & 8.38          & 5.51          & 6.74          & 6.10          \\
   DeepHarmony \cite{deepharmony}           & X         & 0.820         & 0.757         & 0.626         & 0.765             & 5.15          & 7.70          & 6.41          & 10.2           \\ \hline
   \end{tabular}}
   \caption{Evaluation of the methods for two downstream tasks: tissue segmentation and age prediction. Quantitative assessments were performed using a Dice score for the tissue segmentation and a mean absolute error (MAE) value for the age prediction. BlindHarmonyDiff demonstrates overall superior performance in reducing domain gaps in both tasks.}
   \label{table:downmetric}
\end{table*}

\subsection{Downstream task level evaluation}
To evaluate the functional performance of harmonization, we adopted two downstream tasks: brain tissue segmentation and age prediction. Task networks were trained on the target domain dataset, resulting in performance degradation when source domain images were input due to the domain gap. Successful harmonization should bridge this gap, potentially improving downstream task performance. We generated brain tissue masks using FSL FAST \cite{zhang2001segmentation} and obtained age information from the OASIS 3 dataset demographics \cite{lamontagne2019oasis}. For the segmentation network, we employed U-Net \cite{ronneberger2015u}, while a CNN architecture similar to that used in \cite{cole2017predicting} was implemented for age prediction.

Figure \ref{fig:downstream} presents tissue segmentation results using various harmonization methods. BlindHarmonyDiff (column 4) effectively reduces the domain gap between source and target domains, as evidenced by improved tissue segmentation accuracy compared to the non-harmonized case (column 2). Furthermore, BlindHarmonyDiff outperforms other harmonization methods, yielding segmentation results that closely align with the labels.

Table \ref{table:downmetric} presents the quantitative assessment for downstream tasks, including Dice scores for tissue segmentation and mean absolute errors (MAE) for age prediction. BlindHarmonyDiff demonstrates superior Dice scores for tissue segmentation and lower MAE for age prediction compared to other blind harmonization methods, confirming its robust harmonization capabilities. Notably, in tissue segmentation, BlindHarmonyDiff outperforms DeepHarmony, which was trained on each source domain image. This superior performance may be attributed to the blurred output of the U-Net in DeepHarmony.

\subsection{3D vs. 2D in harmonization performance}
\begin{figure}
\setlength{\tabcolsep}{0pt}
\renewcommand{\arraystretch}{0.2}
\centering
   {\tiny
\begin{tabular}{>{\centering\arraybackslash}m{0.02\textwidth}>{\centering\arraybackslash}m{0.11\textwidth}>{\centering\arraybackslash}m{0.11\textwidth}>{\centering\arraybackslash}m{0.11\textwidth}>{\centering\arraybackslash}m{0.11\textwidth}}
        &Source domain & Target domain & BlindHarmonyDiff (3D) &  BlindHarmonyDiff (2D) \\
        &
         \includegraphics[width=0.11\textwidth]{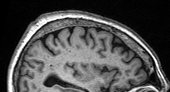}&
         \includegraphics[width=0.11\textwidth]{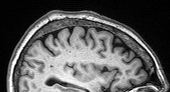}&
         \includegraphics[width=0.11\textwidth]{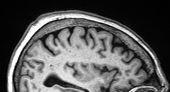}&
         \includegraphics[width=0.11\textwidth]{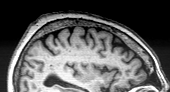}\\
            \rotatebox{90}{\makebox[0.05\textwidth][c]{Zoomed-in}} &
         \includegraphics[width=0.11\textwidth]{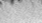}&
         \includegraphics[width=0.11\textwidth]{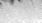}&
         \includegraphics[width=0.11\textwidth]{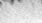}&
         \includegraphics[width=0.11\textwidth]{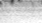}\\
         &
         \includegraphics[width=0.11\textwidth]{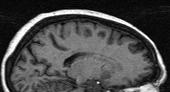}&
         \includegraphics[width=0.11\textwidth]{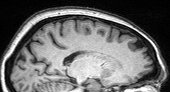}&
         \includegraphics[width=0.11\textwidth]{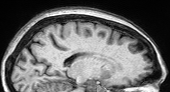}&
         \includegraphics[width=0.11\textwidth]{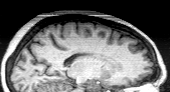}\\
            \rotatebox{90}{\makebox[0.05\textwidth][c]{Zoomed-in}} &
         \includegraphics[width=0.11\textwidth]{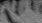}&
         \includegraphics[width=0.11\textwidth]{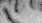}&
         \includegraphics[width=0.11\textwidth]{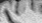}&
         \includegraphics[width=0.11\textwidth]{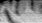}\\
         
     \end{tabular}
     }
     \caption{Comparison of 3D vs. 2D image processing networks. 2D processing shows inter-slice discontinuities (see zoomed-in image), while 3D processing yields homogeneous results across slices.}
   \label{fig:2dvs3d}
 \end{figure}
 
\begin{table*}[]
\centering
      {\footnotesize
   \begin{tabular}{ccccccccc}
   \hline
                         & \multicolumn{2}{c}{Sonata 1.5T}      & \multicolumn{2}{c}{BioGraph 3T}      & \multicolumn{2}{c}{Vision 1.5T} & \multicolumn{2}{c}{Magnetom Vida 3T}\\ 
                         & SSIM($\uparrow$) & PSNR($\uparrow$) & SSIM($\uparrow$) & PSNR($\uparrow$) & SSIM($\uparrow$) & PSNR($\uparrow$) & SSIM($\uparrow$)  & PSNR($\uparrow$) \\ \hline \hline
   BlindHarmonyDiff (3D) & \textbf{0.908} & \textbf{23.4}      & \textbf{0.862} & \textbf{23.3}      & \textbf{0.697} & \textbf{19.3}      & \textbf{0.851} & \textbf{21.3}       \\
   BlindHarmonyDiff (2D) & 0.881          & 21.1               & 0.840          & 21.6               & 0.681          & 18.5               & 0.818          & 19.6                \\ \hline
   \end{tabular}}
   \caption{Quantitative assessment for the comparison between 3D and 2D processing. }
   \label{table:2dvs3d}
\end{table*}

To evaluate the effectiveness of our subject-wise 3D harmonization, we compared the results of 2D and 3D processing. For 2D processing, we trained the 2D rectified flow model at the slice level and calculated the slice-wise normalized cross-correlation in the refinement module.

Figure \ref{fig:2dvs3d} presents exemplary results comparing 3D and 2D processing. The 3D approach yields more homogeneous results, effectively eliminating inter-slice discontinuities present in 2D processing (See zoomed-in image). Quantitative analysis (Table \ref{table:2dvs3d}) further confirms the superior performance of 3D processing over 2D. This superiority may be attributed to the 3D rectified flow's ability to learn inter-slice relationships and the refinement module's consideration of whole-brain correlation information.


\subsection{Refinement module test and Ablation study}
\begin{figure}
\setlength{\tabcolsep}{0pt}
\renewcommand{\arraystretch}{1}
\centering
   {\tiny
\begin{tabular}{>{\centering\arraybackslash}m{0.095\textwidth}>{\centering\arraybackslash}m{0.095\textwidth}>{\centering\arraybackslash}m{0.095\textwidth}>{\centering\arraybackslash}m{0.095\textwidth}>{\centering\arraybackslash}m{0.095\textwidth}}
Source domain & Target domain & BlindHarmonyDiff NCC (ours) & BlindHarmonyDiff MI & BlindHarmonyDiff \quad no refinement \\

        \begin{tikzpicture}
        \node[anchor=south west,inner sep=0] (image) at (0,0) {\includegraphics[width=0.095\textwidth]{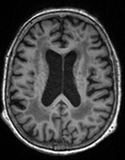}}; 
            \begin{scope}[x={(image.south east)},y={(image.north west)}] 
                \draw[red, very thick, ->] (0.2,0.2) -- (0.3,0.3) ;
            \end{scope}
        \end{tikzpicture}&
        
         \begin{tikzpicture}
        \node[anchor=south west,inner sep=0] (image) at (0,0) {\includegraphics[width=0.095\textwidth]{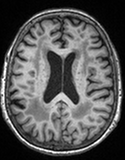}}; 
            \begin{scope}[x={(image.south east)},y={(image.north west)}] 
                \draw[red, very thick, ->] (0.2,0.2) -- (0.3,0.3) ;
            \end{scope}
        \end{tikzpicture}&
        
      \begin{tikzpicture}
        \node[anchor=south west,inner sep=0] (image) at (0,0) {\includegraphics[width=0.095\textwidth]{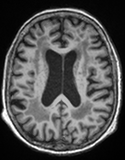}}; 
            \begin{scope}[x={(image.south east)},y={(image.north west)}] 
                \draw[red, very thick, ->] (0.2,0.2) -- (0.3,0.3) ;
            \end{scope}
        \end{tikzpicture}&
         \begin{tikzpicture}
        \node[anchor=south west,inner sep=0] (image) at (0,0) {\includegraphics[width=0.095\textwidth]{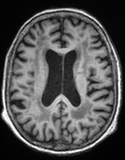}}; 
            \begin{scope}[x={(image.south east)},y={(image.north west)}] 
                \draw[red, very thick, ->] (0.2,0.2) -- (0.3,0.3) ;
            \end{scope}
        \end{tikzpicture}&\begin{tikzpicture}
        \node[anchor=south west,inner sep=0] (image) at (0,0) {\includegraphics[width=0.095\textwidth]{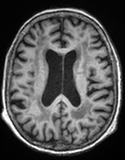}}; 
            \begin{scope}[x={(image.south east)},y={(image.north west)}] 
                \draw[red, very thick, ->] (0.2,0.2) -- (0.3,0.3) ;
            \end{scope}
        \end{tikzpicture}\\
     \end{tabular}
     }
     \caption{Impact of the refinement module on hallucination mitigation. Without the refinement module, BlindHarmonyDiff removes a lesion (fifth column; red arrows), demonstrating hallucination. The refinement module effectively alleviates this issue, (third column; red arrows) producing more reliable images. }
   \label{fig:NCCvsMI}
 \end{figure}
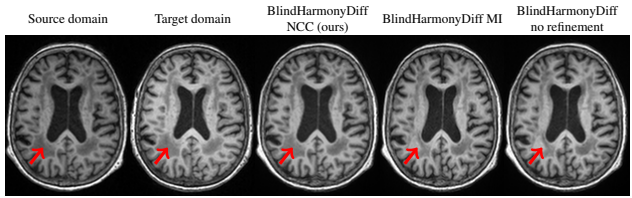
 
\begin{table}[]
\setlength{\tabcolsep}{0.1pt}
\centering
      {\scriptsize
   \begin{tabular}{ccccccccc}
   \hline
Refinement          & \multicolumn{2}{c}{Sonata 1.5T}     & \multicolumn{2}{c}{BioGraph 3T}     & \multicolumn{2}{c}{Vision 1.5T}     & \multicolumn{2}{c}{Magnetom Vida 3T} \\
 module & SSIM($\uparrow$) & PSNR($\uparrow$) & SSIM($\uparrow$) & PSNR($\uparrow$) & SSIM($\uparrow$) & PSNR($\uparrow$) & SSIM($\uparrow$)  & PSNR($\uparrow$) \\ \hline \hline
 NCC               & \textbf{0.908} & \textbf{23.4}      & \textbf{0.862} & \textbf{23.3}      & \textbf{0.697} & \textbf{19.3}      & \textbf{0.851} & \textbf{21.3}       \\
 MI                & 0.877          & 22.6               & 0.836          & 22.4               & 0.693          & 19.2               & 0.827          & 21.1                \\
 None              & 0.870          & 22.3               & 0.827          & 22.1               & 0.689          & 19.1               & 0.820          & 20.9                \\
\hline
   \end{tabular}}
   \caption{Quantitative ablation results show that the full BlindHarmonyDiff with normalized cross-correlation (NCC) outperforms versions with mutual information (MI) or without refinement module.}
   \label{table:ablation}
\end{table}

To assess the contribution of the refinement module, we compared three options: BlindHarmonyDiff with normalized cross-correlation (BlindHarmonyDiff NCC; proposed), BlindHarmonyDiff with mutual information (BlindHarmonyDiff MI), and BlindHarmonyDiff without refinement module (BlindHarmonyDiff no refinement). For BlindHarmonyDiff MI, we updated the harmonized image using gradient ascent with mutual information between histograms of the harmonized and source domain images. We employed kernel density estimation \cite{parzen1962estimation} with a Gaussian kernel (256 bins, $\sigma$ = 0.01) to calculate continuous histogram functions.

The refinement module proves crucial in reducing hallucination and preserving lesion information (Fig. \ref{fig:NCCvsMI}; column 3 vs. column 5). Comparing normalized cross-correlation and mutual information as refinement functions, the former demonstrates superior performance in preserving lesion information (Fig. \ref{fig:NCCvsMI}; column 3 vs. column 4). We believe that it is due to NCC's direct computation in the image domain, as opposed to MI's calculation in the histogram domain. Quantitative analysis (Table \ref{table:ablation}) further confirms that the original BlindHarmonyDiff outperforms variants using mutual information or lacking a refinement module.

For the ablation study results testing the case without edge-to-image model, see Supplementary material \ref{sec:additional}.

\subsection{Contrast conversion test}
To explore the feasibility of applying blind harmonization to contrast conversion (e.g., T$_2$- to T$_1$-weighted images), we applied BlindHarmonyDiff, trained on T$_1$-weighted images, to T$_2$-weighted images. Due to the expected low correlation between T$_1$- and T$_2$-weighted images, we excluded the refinement module in this application.

\begin{figure}
\setlength{\tabcolsep}{0pt}
\renewcommand{\arraystretch}{0}
\centering
   {\scriptsize
\begin{tabular}{>{\centering\arraybackslash}m{0.12\textwidth}>{\centering\arraybackslash}m{0.12\textwidth}>{\centering\arraybackslash}m{0.12\textwidth}>{\centering\arraybackslash}m{0.12\textwidth}}
Source domain & Target domain & BlindHarmonyDiff (ours) & BlindHarmony \\
         \includegraphics[width=0.12\textwidth]{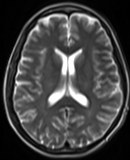}&
         \includegraphics[width=0.12\textwidth]{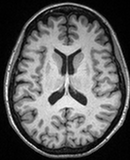}&
         \includegraphics[width=0.12\textwidth]{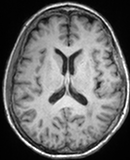}&
         \includegraphics[width=0.12\textwidth]{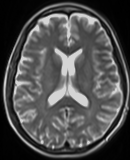}\\
     \end{tabular}
     }
     \caption{Evaluation of BlindHarmonyDiff in bridging a large domain gap (converting T2-weighted image to T1-weighted image) and comparison to the result of the original BlindHarmony.}
   \label{fig:T$_2$}
 \end{figure}

Figure \ref{fig:T$_2$} demonstrates the result of applying BlindHarmonyDiff for T$_2$- to T$_1$-weighted image conversion. BlindHarmonyDiff exhibits capability in handling T$_2$-weighted images as source domain inputs to the previous method. In contrast, BlindHarmony is limited in its ability to perform contrast conversion, which stems from its reliance on an image distance function for optimizing images. Still, the BlindHarmonyDiff results reveal limitations, such as the imperfect conversion of detailed structures. 

\section{Discussion}
\label{sec:discuss}
We introduce BlindHarmonyDiff, a novel 3D blind harmonization framework that leverages an edge-to-image model. Our approach addresses key limitations of existing blind harmonization methods, including inter-slice heterogeneity, moderate image quality, and poor performance for large domain gaps. BlindHarmonyDiff effectively handles images with large domain gaps and generates high-quality 3D harmonized images, as evidenced by higher PSNR and SSIM values and improved performance in downstream tasks compared to previous methods. Furthermore, results indicate that this 3D processing is advantageous for producing high-quality harmonized images by considering inter-slice information. (see Fig. \ref{fig:2dvs3d})

To address the GPU memory challenges associated with training the 3D model, previous methods have employed a patch-based approach \cite{qsmnet, ding2023patched, bieder2023memory, wang2024patch}, which involves partitioning 3D images into smaller patches for neural network training. However, these patch-based methods have not simultaneously achieved efficient memory usage and the ability to capture global context. We propose a multi-stride patch training approach that constructs patches at different strides, enabling the model to incorporate global information while maintaining efficient memory usage. We utilize a normalized index-based method for coordinate encoding. Future research directions may explore alternative spatial encoding techniques, such as sinusoidal encoding, to enhance spatial representation. Additionally, advanced downsampling methods (\eg, Gaussian-weighted non-uniform downsampling) could also be investigated in the context of multi-stride patches. 

As demonstrated in an ablation study, the edge-to-image model alone cannot effectively harmonize source domain images to the target domain. This limitation can be attributed to two main factors. First, although we attempt to generate a homogeneous edge map from the source domain image through threshold tuning, (See Supplementary material \ref{sec:canny}) some edge maps may still exhibit heterogeneous features compared to the edge of the target domain image. Second, the edge-to-image model is susceptible to potential hallucinations \cite{kim2024tackling,aithal2024understanding}. The refinement module plays a crucial role in incorporating source domain structural information into the harmonized image, thereby enabling more accurate and robust harmonization. The erroneous outputs of the edge-to-image model are particularly pronounced when dealing with uncommon features such as lesions and hemorrhages (see Fig. \ref{fig:NCCvsMI}). Consequently, the refinement module is essential for reliable medical image analysis.

Evaluation of harmonization using traveling subjects with image-level correspondence between harmonized and target domain images is a straightforward approach. However, image fidelity metrics such as PSNR and SSIM have known limitations in capturing all aspects of image quality \cite{nilsson2020understanding}. Moreover, the evaluation process with traveling subjects is susceptible to errors stemming from imperfect registration or temporal gaps between acquisitions. Therefore, we also conduct evaluations based on downstream tasks.

In our study, we apply our method to a limited sample of patient data; therefore, further validation with large-scale patient datasets is necessary. Additionally, we select two proof-of-concept downstream tasks that are straightforward to implement. To better analyze the clinical utility of our method, future work should explore its application to a broader range of disease-related downstream tasks.

The application of BlindHarmonyDiff to contrast conversion (\eg, T$_2$-weighted image to T$_1$-weighted image; see Fig. \ref{fig:T$_2$}) produces a T$_1$-weighted image that retains the edges of the input, but the performance remains limited. This limitation arises from the imperfect edge detection for the T$_2$-weighted image, as the threshold tuning was defined based on T$_1$-weighted images. Additionally, we cannot apply the refinement module because the correlation between T$_1$-weighted and T$_2$-weighted images is largely lower than that between two T$_1$-weighted images from different domains. Further research is required to effectively model the relationship between different contrasts.

MRI harmonization shares similarities with domain adaptation in neural networks, as both seek to reduce the domain gap in image data. However, a key distinction lies in their focus: harmonization emphasizes image-level correspondence, while domain adaptation prioritizes the performance of the neural network. MRI presents unique challenges compared to other medical imaging modalities (\eg, CT, X-ray), as achieving a homogeneous dataset without contrast variation is difficult due to long scan times and varying vendors, requiring a harmonization process for large dataset construction. 

Blind harmonization is particularly advantageous when applying harmonization techniques to unknown source domains or those lacking sufficient training data. For instance, when operating an API service that takes MRI inputs, it can be challenging to identify the domain of the input images. In such cases, blind harmonization can enhance the performance of the API service by effectively handling the input data with domain gaps. However, if sufficient source domain data is available for training, employing a harmonization method optimized for a specific source domain  (\eg, DeepHarmony, style transfer; see Fig. \ref{fig:imgeval}) may yield superior results.

\section{Conclusion}
\label{sec:con}
In this study, we proposed BlindHarmonyDiff, a 3D blind harmonization framework leveraging an edge-to-image model to generate high-quality 3D harmonized images with inter-slice consistency. Furthermore, we proposed an efficient 3D patch training scheme with multi-stride patches and a refinement module to effectively suppress hallucination, ensuring practicality and clinical reliability. We have demonstrated excellent performance of our BlindHarmonyDiff in image quality metrics as well as in downstream task-based metrics over prior arts across diverse MR scanners.
{
    \small
    \bibliographystyle{ieeenat_fullname}
    \bibliography{main}
}
\clearpage
\appendix
\renewcommand{\thesection}{S\arabic{section}}
\renewcommand{\thetable}{S\arabic{table}}
\setcounter{table}{0} 
\renewcommand{\thefigure}{S\arabic{figure}}
\setcounter{figure}{0} 
\clearpage
\setcounter{page}{1}
\maketitlesupplementary

\begin{table*}[!htbp]
   \begin{center}
      {\footnotesize
   \begin{tabular}{cccccc}
   \hline
              & Target domain & \multicolumn{4}{c}{Source domain}  \\
              \hline
   Manufacturer                 & Siemens                   & Siemens                       & Siemens                       & Siemens                       & Siemens                       \\
   Scanner version              & TIM Trio                  & Sonata                        & BioGraph mMR                  & Vision                        & Magnetom Vida       \\
   Magnetic field strength (T)  & 3                         & 1.5                           & 3                             & 1.5                             & 3                           \\
   Matrix size                  & 176$\times$256$\times$256 & 160$\times$256$\times$256     & 176$\times$240$\times$256     & 128$\times$256$\times$256     & 176$\times$240$\times$256     \\
   Resolution ($mm^3$)          & 1$\times$1$\times$1       & 1$\times$1$\times$1           & 1.2$\times$1.05$\times$1.05   & 1.25$\times$1$\times$1        & 1.2$\times$1.05$\times$1.05   \\
   TR/TI (s)                    & 2.4/1                     & 1.9/1.1                       & 2.3/0.9                       & 9.7/unknown                   & 2.3/0.9                       \\
   TE (ms)                      & 3.16                      & 3.93                          & 2.95                          & 4.0                           & 2.95                          \\
   Flip angle ($^{\circ}$)      & 8                         & 15                            & 9                             & 10                            & 9                             \\
   Paired data (train/val)      & -                         & 20/1                          & 190/7                         & 67/6                          & 47/4                          \\
   Test subject (image-level)   & -                         & 7                             & 30                            & 12                            & 10                            \\
   Test subject (downstream)    & -                         & 8                             & 134                           & 25                            & 54                            \\
              \hline
   \end{tabular}
   }
\end{center}
   \caption{Representative scan parameters for various domains in the OASIS3 dataset are presented. A single data domain may encompass multiple parameter configurations. The `Paired data' row indicates the size of the dataset used for training DeepHarmony and the style transfer network.  }
   \label{table:dataset}
\end{table*}

\begin{table*}[!htbp]
\centering
      {\footnotesize
   \begin{tabular}{ccccccccccc}
   \hline
    &       & \multicolumn{2}{c}{Sonata 1.5T}     & \multicolumn{2}{c}{BioGraph 3T}     & \multicolumn{2}{c}{Vision 1.5T}     & \multicolumn{2}{c}{Magnetom Vida 3T} \\ 
    & Blind & SSIM($\uparrow$) & PSNR($\uparrow$) & SSIM($\uparrow$) & PSNR($\uparrow$) & SSIM($\uparrow$) & PSNR($\uparrow$) & SSIM($\uparrow$)  & PSNR($\uparrow$) \\ \hline \hline
   Source domain                            &      & 0.795          & 17.3               & 0.808          & 18.2               & 0.534          & 12.2               & 0.780          & 16.0                \\ \hline
   BlindHarmonyDiff (ours)                  & O    & \textbf{0.873} & \textbf{21.2}      & \textbf{0.819} & \textbf{20.9}      & \textbf{0.611} & \textbf{16.5}      & \textbf{0.814} & \textbf{19.5}       \\
   BlindHarmony \cite{jeong2023blindharmony} & O    & 0.782          & 17.4               & 0.810          & 18.9               & 0.578          & 14.3               & 0.783          & 16.5                \\
   SSIMH \cite{guan2022fast}                 & O    & 0.734          & 13.5               & 0.701          & 14.2               & 0.496          & 12.6               & 0.709          & 13.7                \\ 
   Histogram matching                       & O    & 0.765          & 14.5               & 0.780          & 16.0               & 0.566          & 14.0               & 0.772          & 15.4                \\ \hline
   \end{tabular}}
   \caption{Quantitative assessment using PSNR and SSIM metrics in 2D processing case, calculated relative to the target domain images.}
   \label{table:2dmetric}
\end{table*}

\begin{table*}[!htbp]
\centering
      {\footnotesize
   \begin{tabular}{ccccccccc}
   \hline
                        & \multicolumn{2}{c}{Sonata 1.5T}     & \multicolumn{2}{c}{BioGraph 3T}     & \multicolumn{2}{c}{Vision 1.5T}     & \multicolumn{2}{c}{Magnetom Vida 3T} \\
 Edge-to-image model    & SSIM($\uparrow$) & PSNR($\uparrow$) & SSIM($\uparrow$) & PSNR($\uparrow$) & SSIM($\uparrow$) & PSNR($\uparrow$) & SSIM($\uparrow$)  & PSNR($\uparrow$) \\ \hline \hline
 Conditional                      & \textbf{0.908} & \textbf{23.4}      & \textbf{0.862} & \textbf{23.3}      & \textbf{0.697} & 19.3      & \textbf{0.851} & \textbf{21.3}       \\
 Unconditional            & 0.830          & 21.6               & 0.765          & 21.5               & 0.640          & 19.3               & 0.770          & 20.4                \\
 None           & 0.769          & 19.2               & 0.764          & 20.7               & 0.630          & 18.9               & 0.750          & 19.7                \\
\hline
   \end{tabular}}
   \caption{Quantitative ablation studies demonstrate the superiority of the full BlindHarmonyDiff framework incorporating the conditional edge-to-image model. This configuration consistently outperforms the case employing the unconditional edge-to-image model or the case without the edge-to-image model. }
   \label{table:ablatione2i}
\end{table*}

\section{Threshold of Canny edge detction}
\label{sec:canny}
\begin{figure}
 \setlength{\tabcolsep}{0pt}
\renewcommand{\arraystretch}{0}
\centering
   {\scriptsize
\begin{tabular}{>{\centering\arraybackslash}m{0.1\textwidth}>{\centering\arraybackslash}m{0.08\textwidth}>{\centering\arraybackslash}m{0.08\textwidth}}
(a) & Domain 1 & Domain 2  \\ [2pt]
        Fixed edge percentage &
         \includegraphics[width=0.08\textwidth]{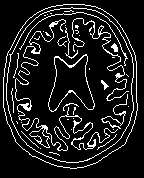}&
         \includegraphics[width=0.08\textwidth]{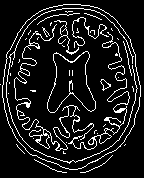}\\
        Fixed threshold &
         \includegraphics[width=0.08\textwidth]{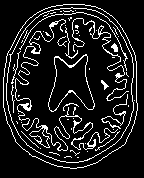}&
         \includegraphics[width=0.08\textwidth]{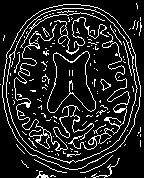}\\
     \end{tabular}
     
\begin{tabular}{>{\centering\arraybackslash}m{0.02\textwidth}>{\centering\arraybackslash}m{0.078\textwidth}>{\centering\arraybackslash}m{0.078\textwidth}>{\centering\arraybackslash}m{0.078\textwidth}>{\centering\arraybackslash}m{0.078\textwidth}}
 (b) &Image & 8\% & 10\% & 6\%\\ [2pt]
 &
         \includegraphics[width=.078\textwidth]{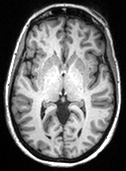}&
         \includegraphics[width=.078\textwidth]{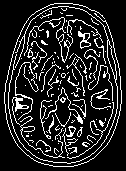}&
         \includegraphics[width=.078\textwidth]{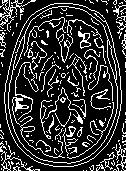}&
         \includegraphics[width=.078\textwidth]{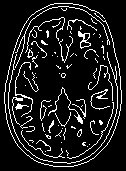}\\
     \end{tabular}
     }
     \caption{Threshold test}
   \label{fig:per}
 \end{figure}

Canny edge detection can be parameter-sensitive. However, we chose this method for its easy implementation and reproducibility across various domains. To address the parameter sensitivity, we employ a fixed edge percentage approach rather than a fixed threshold. Figure \ref{fig:per}a demonstrates that using a fixed percentage yields more consistent edge maps across different domains than a fixed threshold does. 

We empirically determined 8\% edge, because it provides good edge detection for brain images, as illustrated in the Figure \ref{fig:per}b. Although this threshold selection is somewhat heuristic, it produces reliable edge maps for brain structures across our dataset. We recognize that this approach might occasionally result in inconsistent edge detection and produce poor edge-to-image model output (even inferior to the one without harmonization). However, as noted in the Discussion, the refinement module is designed to mitigate such inconsistencies.

\section{Dataset description}
\label{sec:data}
The BlindHarmonyDiff framework presented in this study was trained and evaluated using the OASIS3 dataset \cite{lamontagne2019oasis}, with a target domain comprising images acquired on a Siemens TIM Trio 3T MR scanner. For the source domain, images from four additional scanners were employed: Siemens Sonata 1.5T, Siemens BioGraph mMR PET-MR 3T, Siemens Vision 1.5T, and Siemens Magnetom Vida 3T MR. All images underwent resampling to a uniform resolution of 1.2$\times$1.2$\times$1.25 mm (matrix size: 144$\times$208$\times$92) and were subjected to percentile-based min-max normalization (1st-99th) at the subject level. Table \ref{table:dataset} presents detailed acquisition parameters for each scanner. For target domain training, encompassing edge-to-image model training and downstream task networks training, the dataset was partitioned into train/validation/test sets comprising 717, 29, and 135 subjects, respectively.

\section{Implementation details}
\label{sec:implementation}
\subsection{Edge-to-image model}
The 3D edge-to-image model was trained on cubic patches of size 64$\times$64$\times$64 with a batch size of 8, using 2 NVIDIA Quadro RTX 8000 GPUs. The model was optimized using the Adam optimizer with an L2 loss function and a fixed learning rate of 0.00005 for 150,000 steps.

The architecture is based on a 3D U-Net \cite{ronneberger2015u}, with five input channels: a grayscale noise channel, a conditional grayscale edge channel, and three positional coordinate channels. The output is a single grayscale channel. The first layer has 64 channels, with depth multipliers of \{1,\,2,\,4,\,8\} applied at successive layers. Each block includes two 3D convolution layers, RMS normalization, and SiLU activation. Downsampling is performed via resampling, where spatial blocks are reorganized into the channel dimension, while upsampling combines nearest-neighbor interpolation and convolution. Skip connections transfer encoder features directly to the decoder, preserving spatial details.

Data augmentation includes random flipping along the x, y, and z axes to increase the diversity of the training data. Multi-stride patches were applied during training to capture multi-scale features, enabling the model to effectively learn both local and global contexts. 

During the sampling process, the model used a step size of 16. The ordinary differential equation solver employed the midpoint method, with both absolute and relative error tolerances set to 0.00005.
   
\subsection{DeepHarmony}
The 3D U-Net \cite{ronneberger2015u} architecture was employed for the implementation of DeepHarmony \cite{deepharmony}. The network comprises 4 Down blocks and 4 Up blocks, with each block consisting of two sequential combinations of a convolution layer (5$\times$5$\times$5), batch normalization, and leaky-ReLU activation. Maxpooling is utilized in the Down blocks, while transpose convolution is employed in the Up blocks. The U-Net also incorporates skip connections. Training was conducted using L1 loss and the SGD optimizer with a learning rate of 0.001. The checkpoint exhibiting the best validation loss was selected for evaluation. The dataset size varies by domain, as detailed in Table \ref{table:dataset}.

\subsection{Style transfer}
The style transfer network employed a 2D paired CycleGAN architecture \cite{zhu2017unpaired}. The generator network comprises 2 convolutional layers with instance normalization and ReLU activation, followed by 9 residual blocks, and concludes with 3 convolutional layers with instance normalization and ReLU activation. Each residual block incorporates a residual connection consisting of 2 convolutional layers with instance normalization and ReLU activation. The discriminator is composed of 5 convolutional layers, 4 leaky ReLU activations, and 4 instance normalization layers. The CycleGAN training objective incorporates identity loss, cycle consistency loss, and adversarial loss. The training was conducted for a maximum of 200 epochs using the Adam optimizer with a learning rate of 0.0002. The dataset size varies by domain, as detailed in Table \ref{table:dataset}.

\subsection{BlindHarmony}
For training the normalizing flow, we adopted the Neural Spline Flow (NSF) architecture \cite{durkan2019neural} with rational quadratic (RQ) spline coupling layers. The majority of hyperparameters were set to match those used in the original NSF paper for ImageNet experiments. Specifically, we set the tail bound B to 3 and the number of bins K for RQ spline coupling layers to 8. We implemented a multiscale architecture akin to Glow, with each network layer comprising 7 transformation steps: an actnorm layer, an invertible 1$\times$1 convolution, an RQ spline coupling transform, and another 1$\times$1 convolution. The network consists of 4 layers, resulting in 28 coupling transformation steps. The subnetworks parameterizing the RQ splines incorporate 3 residual blocks and batch normalization layers. We optimized the parameters using an Adam optimizer with an initial learning rate of 0.0005 and cosine annealing, iterating up to 10K steps.

\subsection{Tissue segmentation network}
The 2D U-Net \cite{ronneberger2015u}  architecture was employed for implementing the tissue segmentation network. The network comprises 4 Down blocks and 4 Up blocks. Each Down block consists of a maxpooling layer (2$\times$2), followed by two sequences of convolution layer (3$\times$3), batch normalization, and ReLU activation. Up blocks are composed of a bilinear upsampling layer (2$\times$2), followed by two sequences of convolution layer (3$\times$3), batch normalization, and ReLU activation. The U-Net incorporates skip connections between corresponding Down and Up blocks. The training was conducted using L1 loss and the Adam optimizer with a learning rate of 0.0001. The checkpoint exhibiting the best validation loss was selected for evaluation.

\subsection{Age prediction network}
A 3D CNN architecture similar to that proposed in \cite{cole2017predicting} was implemented for age prediction. The network comprises 5 repeated blocks, each consisting of a (3$\times$3$\times$3) convolutional layer with stride 1, followed by ReLU activation, another (3$\times$3$\times$3) convolutional layer with stride 1, a 3D batch-normalization layer, ReLU activation, and a (2$\times$2$\times$2) max-pooling layer with stride 2. The number of feature channels is initialized at eight in the first block and doubles after each max-pooling operation to derive a rich representation of the brain structure. The final age prediction is obtained through a fully connected layer, mapping the output of the last block to a single value. Training was conducted using L1 loss and the SGD optimizer with a learning rate of 0.0001. The model checkpoint exhibiting the best validation loss was selected for evaluation.

\section{Addtional test}
\label{sec:additional}
\begin{table}[]
\centering
      {\scriptsize
   \begin{tabular}{ccccc}
   \hline
                            & \multicolumn{2}{c}{2D}                & \multicolumn{2}{c}{3D}                \\
                            & SSIM($\uparrow$)  & PSNR($\uparrow$)  & SSIM($\uparrow$)  & PSNR($\uparrow$)  \\ \hline \hline
    Whole image             & 0.813             & 24.0              & -                 & -                 \\ \hline
    Multi-stride patch  & \textbf{0.817}             & \textbf{24.1}              & \textbf{0.920}             & \textbf{27.5}              \\
    Simple patch & 0.809             & 23.9              & 0.905             & 26.9              \\
\hline
   \end{tabular}}
   \caption{Comparison between multi-stride patch training and simple patch training.}
   \label{table:multiresol}
\end{table}

\subsection{Evaluation of multi-stride patch in 2D}
To assess the effectiveness of multi-stride patch training, we conducted experiments in both 2D and 3D scenarios. For 2D cases, we compared whole image training, multi-stride patch training, and simple patch training. In 3D cases, we evaluated multi-stride and simple patch training. PSNR and SSIM metrics were calculated relative to the original image to quantify performance.

Our experimental results (Table \ref{table:multiresol}) demonstrate that multi-stride patch training consistently outperforms simple patch training in both 2D and 3D scenarios. Notably, in 2D cases, multi-stride patch training achieved higher PSNR and SSIM values than even whole-image training. These results highlight the ability of multi-stride patch training to enhance image quality and structural similarity, underscoring its effectiveness.

\subsection{Evaluation of harmonization in 2D}
We compared the result when the processing is based on 2D (\ie, slice-wise calculation). BlindHarmonyDiff was implemented using a 2D rectified flow model trained at the slice level, without utilizing patch-based training. The refinement module employed slice-wise normalized cross-correlation calculations. For SSIMH \cite{guan2022fast}, the approach involved averaging whole slices of target domain images and replacing the low-frequency components of source domain images with these averaged target domain images using a 2D discrete cosine transform. Histogram matching was performed by storing the histogram of target domain images and matching the source domain image histogram to the target domain histogram at the slice level. Metrics were calculated slice-wise, while the 2D metrics in Table \ref{table:2dvs3d} were computed for the whole 3D brain volume.

Quantitative analysis results  (Table \ref{table:2dmetric}) demonstrate that the overall performance trends are consistent with the 3D case. BlindHarmonyDiff outperforms other blind harmonization techniques, including BlindHarmony \cite{jeong2023blindharmony}, histogram matching, and SSIMH \cite{guan2022fast}, in both PSNR and SSIM metrics. Notably, the normalizing flow-based BlindHarmony method showed superior performance compared to SSIMH and histogram matching, which aligns with findings reported by previous study \cite{jeong2023blindharmony}.

\subsection{Ablation test: the case without edge-to-image model}
To evaluate the contribution of the edge-to-image model in harmonization, we conducted a comparative analysis of three configurations: BlindHarmonyDiff with the proposed conditional edge-to-image model, BlindHarmonyDiff with an unconditional edge-to-image model, and BlindHarmonyDiff without an edge-to-image model. For the unconditional edge-to-image model variant, we did not train a separate model but instead input random edge conditions when generating input for the refinement module. In the case without an edge-to-image model, we applied only the refinement module to the averaged image from the target domain training dataset.

Quantitative analysis (Table \ref{table:ablatione2i}) demonstrates that the original BlindHarmonyDiff, incorporating the conditional edge-to-image model, outperforms both the unconditional variant and the version without an edge-to-image model. While the refinement module does enforce some structural information from the source domain image, the initial starting point provided by the edge-to-image model proves crucial for achieving high-quality harmonization. These results underscore the importance of the conditional edge-to-image model in the BlindHarmonyDiff framework for effective image harmonization.

\section{Visual examples}
\begin{figure*}
\setlength{\tabcolsep}{0pt}
\renewcommand{\arraystretch}{0}
\centering
   {\scriptsize
\begin{tabular}{>{\centering\arraybackslash}m{0.105\textwidth}>{\centering\arraybackslash}m{0.105\textwidth}>{\centering\arraybackslash}m{0.105\textwidth}>{\centering\arraybackslash}m{0.105\textwidth}>{\centering\arraybackslash}m{0.105\textwidth}>{\centering\arraybackslash}m{0.105\textwidth}>{\centering\arraybackslash}m{0.105\textwidth}>{\centering\arraybackslash}m{0.105\textwidth}>{\centering\arraybackslash}m{0.105\textwidth}}
&Source domain & Target domain & BlindHarmonyDiff (ours) & BlindHarmony \cite{jeong2023blindharmony} & SSIMH \cite{guan2022fast}& Histogram matching  & Style transfer \cite{cyclegan1,cyclegan2} & DeepHarmony \cite{deepharmony}\\

          Sonata 1.5T&
         \includegraphics[width=0.105\textwidth]{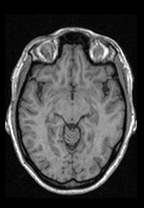}&
         \includegraphics[width=0.105\textwidth]{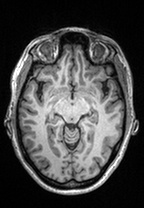}&
         \includegraphics[width=0.105\textwidth]{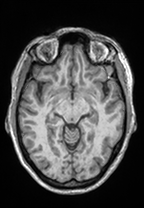}&
         \includegraphics[width=0.105\textwidth]{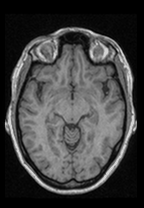}&
         \includegraphics[width=0.105\textwidth]{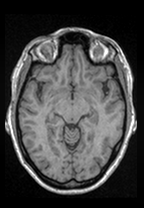}&
         \includegraphics[width=0.105\textwidth]{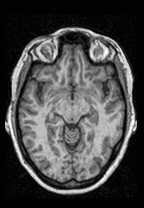}&
         \includegraphics[width=0.105\textwidth]{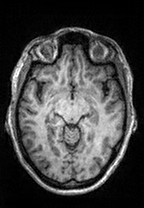}&
         \includegraphics[width=0.105\textwidth]{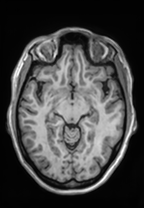}\\
           Sonata 1.5T&
         \includegraphics[width=0.105\textwidth]{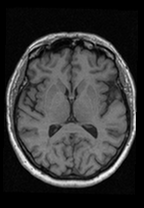}&
         \includegraphics[width=0.105\textwidth]{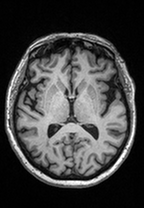}&
         \includegraphics[width=0.105\textwidth]{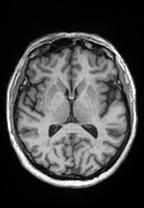}&
         \includegraphics[width=0.105\textwidth]{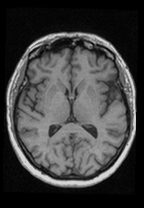}&
         \includegraphics[width=0.105\textwidth]{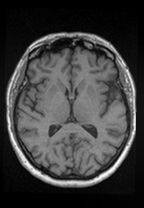}&
         \includegraphics[width=0.105\textwidth]{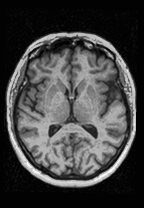}&
         \includegraphics[width=0.105\textwidth]{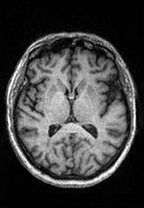}&
         \includegraphics[width=0.105\textwidth]{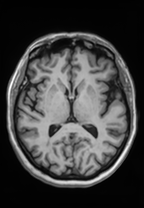}\\
         BioGraph 3T&
         \includegraphics[width=0.105\textwidth]{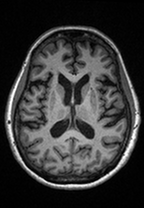}&
         \includegraphics[width=0.105\textwidth]{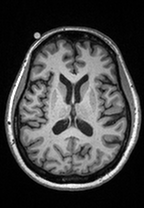}&
         \includegraphics[width=0.105\textwidth]{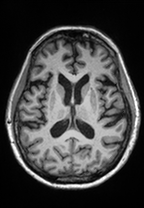}&
         \includegraphics[width=0.105\textwidth]{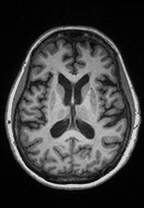}&
         \includegraphics[width=0.105\textwidth]{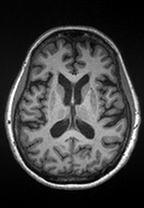}&
         \includegraphics[width=0.105\textwidth]{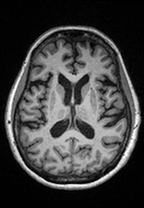}&
         \includegraphics[width=0.105\textwidth]{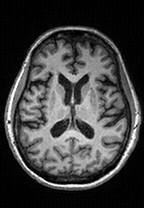}&
         \includegraphics[width=0.105\textwidth]{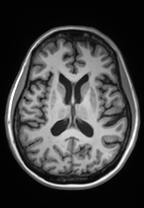}\\
         BioGraph 3T&
         \includegraphics[width=0.105\textwidth]{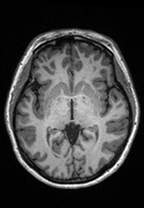}&
         \includegraphics[width=0.105\textwidth]{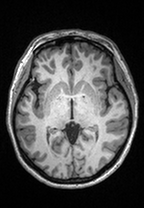}&
         \includegraphics[width=0.105\textwidth]{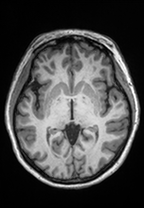}&
         \includegraphics[width=0.105\textwidth]{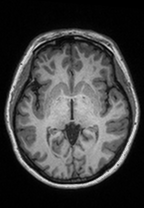}&
         \includegraphics[width=0.105\textwidth]{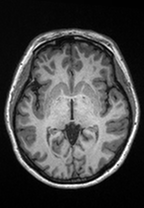}&
         \includegraphics[width=0.105\textwidth]{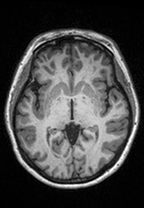}&
         \includegraphics[width=0.105\textwidth]{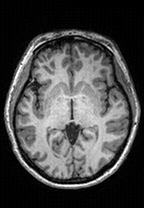}&
         \includegraphics[width=0.105\textwidth]{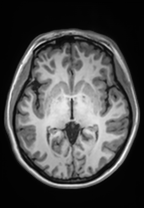}\\
         
        Vision 1.5T&
         \includegraphics[width=0.105\textwidth]{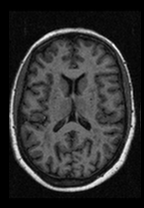}&
         \includegraphics[width=0.105\textwidth]{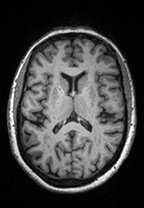}&
         \includegraphics[width=0.105\textwidth]{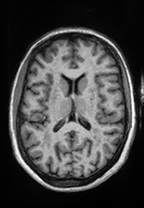}&
         \includegraphics[width=0.105\textwidth]{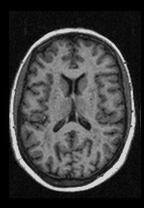}&
         \includegraphics[width=0.105\textwidth]{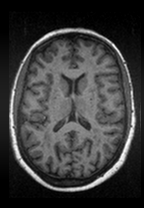}&
         \includegraphics[width=0.105\textwidth]{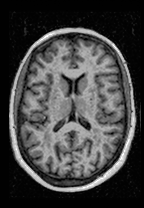}&
         \includegraphics[width=0.105\textwidth]{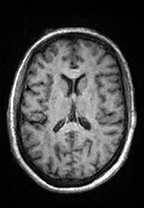}&
         \includegraphics[width=0.105\textwidth]{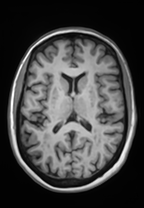}\\
        Vision 1.5T&
         \includegraphics[width=0.105\textwidth]{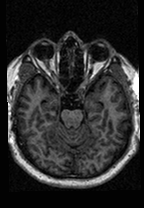}&
         \includegraphics[width=0.105\textwidth]{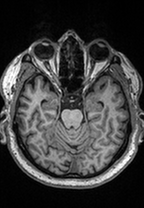}&
         \includegraphics[width=0.105\textwidth]{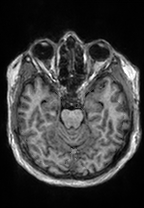}&
         \includegraphics[width=0.105\textwidth]{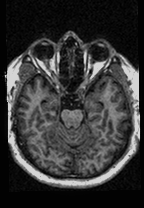}&
         \includegraphics[width=0.105\textwidth]{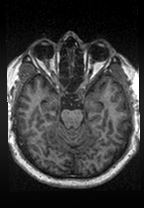}&
         \includegraphics[width=0.105\textwidth]{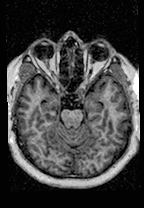}&
         \includegraphics[width=0.105\textwidth]{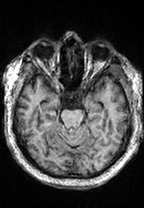}&
         \includegraphics[width=0.105\textwidth]{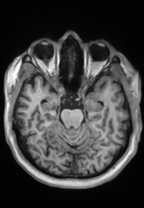}\\
         
         Magnetom Vida 3T&
         \includegraphics[width=0.105\textwidth]{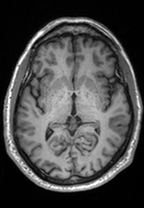}&
         \includegraphics[width=0.105\textwidth]{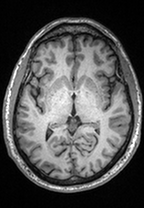}&
         \includegraphics[width=0.105\textwidth]{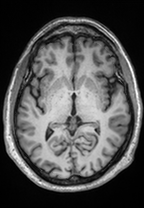}&
         \includegraphics[width=0.105\textwidth]{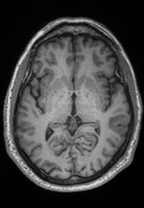}&
         \includegraphics[width=0.105\textwidth]{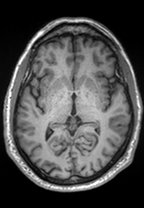}&
         \includegraphics[width=0.105\textwidth]{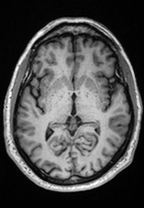}&
         \includegraphics[width=0.105\textwidth]{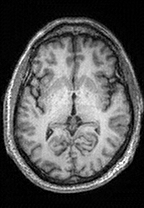}&
         \includegraphics[width=0.105\textwidth]{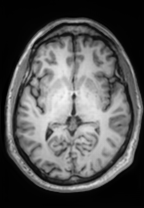}\\
         Magnetom Vida 3T&
         \includegraphics[width=0.105\textwidth]{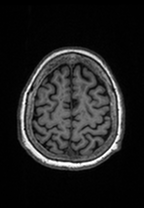}&
         \includegraphics[width=0.105\textwidth]{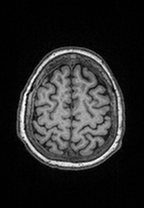}&
         \includegraphics[width=0.105\textwidth]{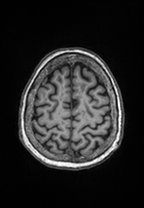}&
         \includegraphics[width=0.105\textwidth]{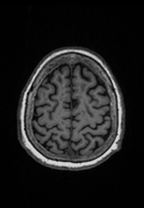}&
         \includegraphics[width=0.105\textwidth]{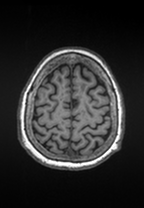}&
         \includegraphics[width=0.105\textwidth]{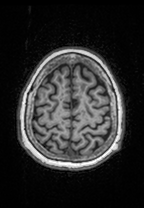}&
         \includegraphics[width=0.105\textwidth]{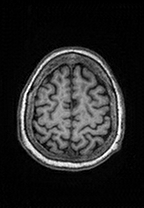}&
         \includegraphics[width=0.105\textwidth]{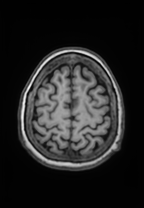}\\
     \end{tabular}
     }
     \caption{Visual comparison of harmonization results on traveling subjects, demonstrating the effectiveness of BlindHarmonyDiff against the other blind harmonization methods.}
   \label{fig:imgsupp}
 \end{figure*}

\begin{figure*}
\setlength{\tabcolsep}{0pt}
\renewcommand{\arraystretch}{0}
\centering
   {\scriptsize
\begin{tabular}{>{\centering\arraybackslash}m{0.1\textwidth}>{\centering\arraybackslash}m{0.1\textwidth}>{\centering\arraybackslash}m{0.1\textwidth}>{\centering\arraybackslash}m{0.1\textwidth}>{\centering\arraybackslash}m{0.1\textwidth}>{\centering\arraybackslash}m{0.1\textwidth}>{\centering\arraybackslash}m{0.1\textwidth}>{\centering\arraybackslash}m{0.1\textwidth}>{\centering\arraybackslash}m{0.1\textwidth}>{\centering\arraybackslash}m{0.1\textwidth}}
& Source domain image &No harmonization & Label & BlindHarmonyDiff (ours) & BlindHarmony \cite{jeong2023blindharmony}   & SSIMH \cite{guan2022fast}& Histogram matching  & Style transfer \cite{cyclegan1,cyclegan2} & DeepHarmony \cite{deepharmony}\\
         Sonata 1.5T&
         \includegraphics[width=0.100\textwidth]{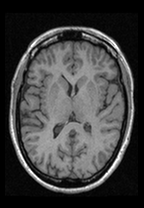}&
         \includegraphics[width=0.100\textwidth]{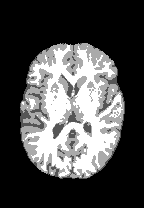}&
         \includegraphics[width=0.100\textwidth]{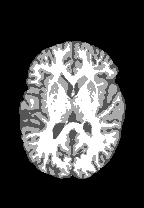}&
         \includegraphics[width=0.100\textwidth]{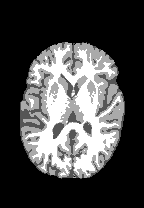}&
         \includegraphics[width=0.100\textwidth]{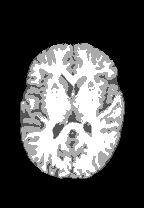}&
         \includegraphics[width=0.100\textwidth]{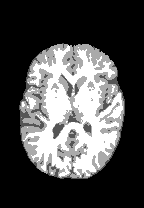}&
         \includegraphics[width=0.100\textwidth]{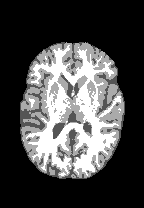}&
         \includegraphics[width=0.100\textwidth]{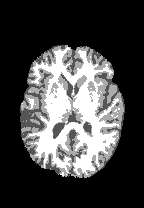}&
         \includegraphics[width=0.100\textwidth]{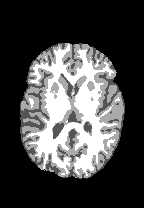}\\
         Sonata 1.5T&
         \includegraphics[width=0.100\textwidth]{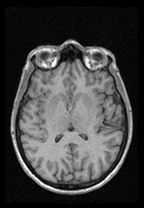}&
         \includegraphics[width=0.100\textwidth]{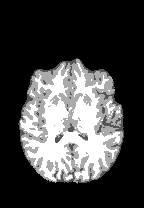}&
         \includegraphics[width=0.100\textwidth]{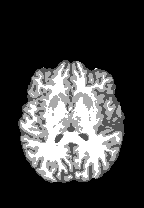}&
         \includegraphics[width=0.100\textwidth]{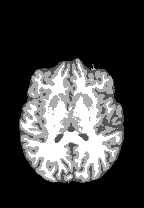}&
         \includegraphics[width=0.100\textwidth]{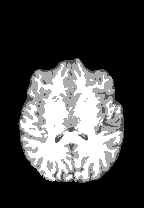}&
         \includegraphics[width=0.100\textwidth]{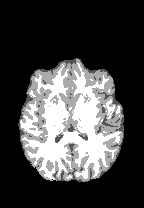}&
         \includegraphics[width=0.100\textwidth]{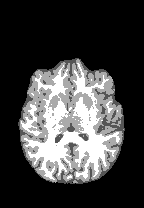}&
         \includegraphics[width=0.100\textwidth]{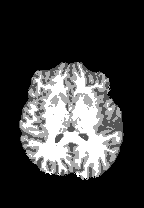}&
         \includegraphics[width=0.100\textwidth]{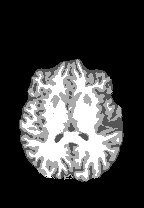}\\
         
         BioGraph 3T&
         \includegraphics[width=0.100\textwidth]{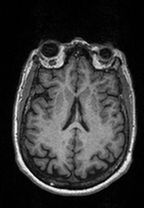}&
         \includegraphics[width=0.100\textwidth]{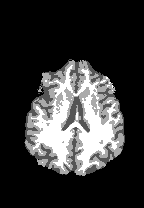}&
         \includegraphics[width=0.100\textwidth]{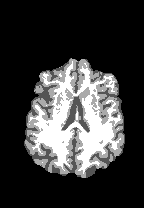}&
         \includegraphics[width=0.100\textwidth]{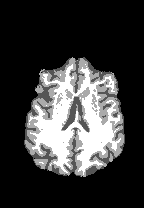}&
         \includegraphics[width=0.100\textwidth]{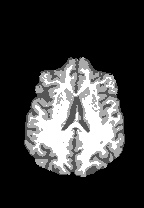}&
         \includegraphics[width=0.100\textwidth]{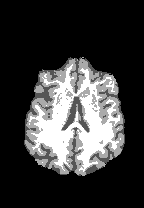}&
         \includegraphics[width=0.100\textwidth]{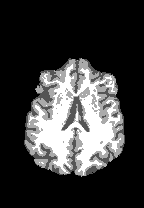}&
         \includegraphics[width=0.100\textwidth]{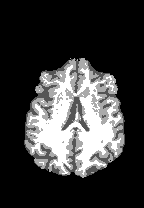}&
         \includegraphics[width=0.100\textwidth]{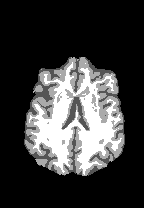}\\
         BioGraph 3T&
         \includegraphics[width=0.100\textwidth]{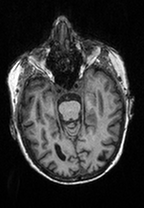}&
         \includegraphics[width=0.100\textwidth]{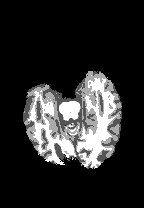}&
         \includegraphics[width=0.100\textwidth]{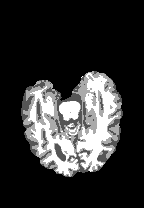}&
         \includegraphics[width=0.100\textwidth]{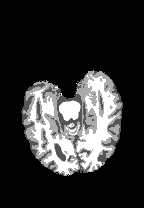}&
         \includegraphics[width=0.100\textwidth]{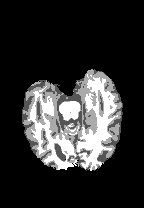}&
         \includegraphics[width=0.100\textwidth]{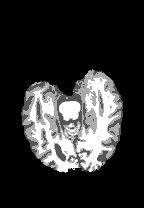}&
         \includegraphics[width=0.100\textwidth]{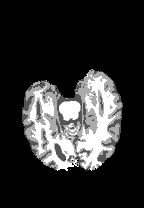}&
         \includegraphics[width=0.100\textwidth]{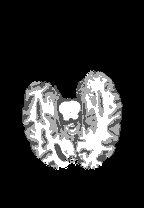}&
         \includegraphics[width=0.100\textwidth]{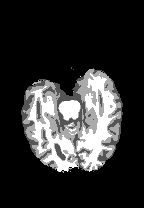}\\
         
        Vision 1.5T&
         \includegraphics[width=0.100\textwidth]{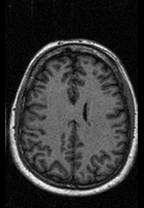}&
         \includegraphics[width=0.100\textwidth]{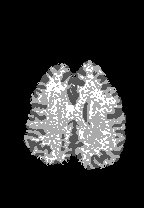}&
         \includegraphics[width=0.100\textwidth]{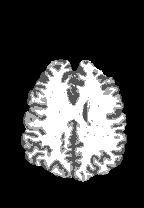}&
         \includegraphics[width=0.100\textwidth]{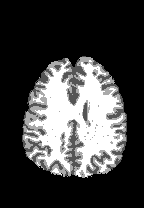}&
         \includegraphics[width=0.100\textwidth]{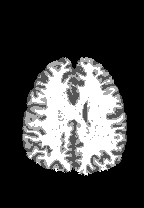}&
         \includegraphics[width=0.100\textwidth]{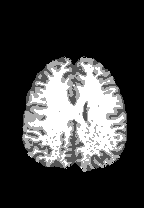}&
         \includegraphics[width=0.100\textwidth]{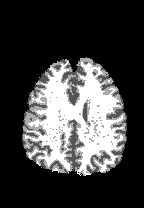}&
         \includegraphics[width=0.100\textwidth]{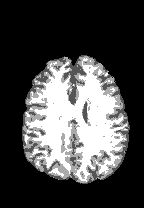}&
         \includegraphics[width=0.100\textwidth]{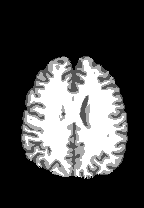}\\
        Vision 1.5T&
         \includegraphics[width=0.100\textwidth]{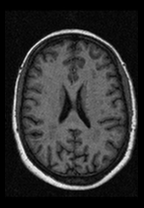}&
         \includegraphics[width=0.100\textwidth]{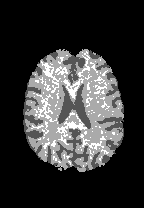}&
         \includegraphics[width=0.100\textwidth]{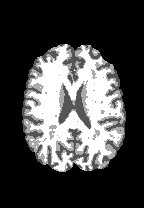}&
         \includegraphics[width=0.100\textwidth]{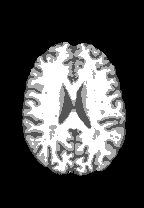}&
         \includegraphics[width=0.100\textwidth]{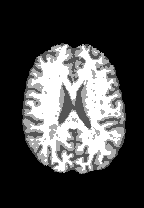}&
         \includegraphics[width=0.100\textwidth]{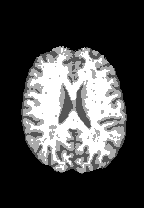}&
         \includegraphics[width=0.100\textwidth]{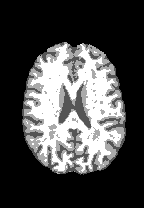}&
         \includegraphics[width=0.100\textwidth]{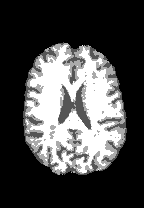}&
         \includegraphics[width=0.100\textwidth]{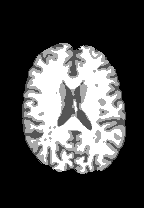}\\
         
         Magnetom Vida 3T&
         \includegraphics[width=0.100\textwidth]{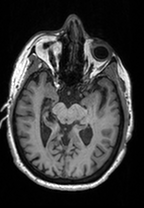}&
         \includegraphics[width=0.100\textwidth]{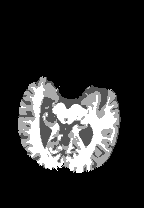}&
         \includegraphics[width=0.100\textwidth]{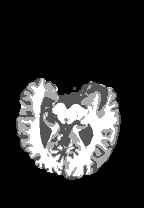}&
         \includegraphics[width=0.100\textwidth]{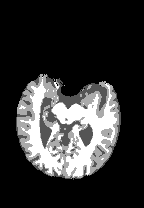}&
         \includegraphics[width=0.100\textwidth]{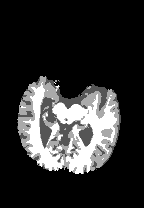}&
         \includegraphics[width=0.100\textwidth]{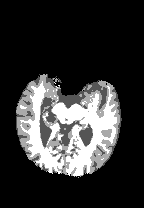}&
         \includegraphics[width=0.100\textwidth]{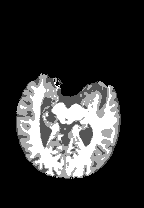}&
         \includegraphics[width=0.100\textwidth]{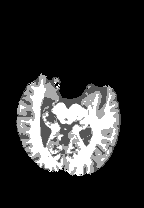}&
         \includegraphics[width=0.100\textwidth]{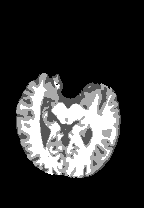}\\
         Magnetom Vida 3T&
         \includegraphics[width=0.100\textwidth]{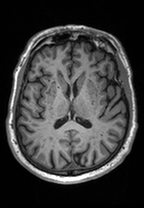}&
         \includegraphics[width=0.100\textwidth]{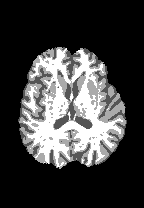}&
         \includegraphics[width=0.100\textwidth]{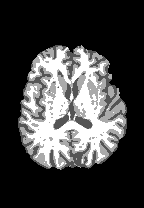}&
         \includegraphics[width=0.100\textwidth]{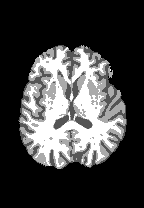}&
         \includegraphics[width=0.100\textwidth]{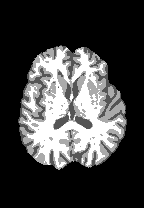}&
         \includegraphics[width=0.100\textwidth]{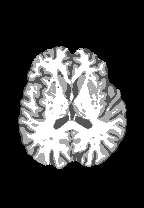}&
         \includegraphics[width=0.100\textwidth]{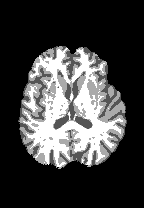}&
         \includegraphics[width=0.100\textwidth]{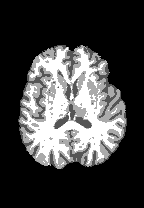}&
         \includegraphics[width=0.100\textwidth]{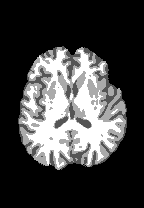}\\
     \end{tabular}
     }
     \caption{Tissue segmentation results for the various harmonization methods, with the segmentation network trained on the target domain data.  Each column represents a different harmonization approach applied to the input images.}
   \label{fig:segsupp}
 \end{figure*}

Exemplary images of the image level evaluation and the downstream task evaluation are illustrated in Figures \ref{fig:imgsupp} and \ref{fig:segsupp}.

\end{document}